# Distinguishing the Topological Charge of Vortex Beam via Fourier Back Plane Imaging with Chiral Gammadion Structure

Yangzhe Guo[1], Jing Li[2] and Yurui Fang[1,*]

[1.] *School of Physics, Dalian University of Technology, Dalian 116024, P.R. China.*

[2.] *Key Laboratory of Photochemical Conversion and Optoelectronic Materials, Technical Institute of Physics and Chemistry, Chinese Academy of Sciences, Beijing 100190, China*

*Corresponding author: Yurui Fang (yrfang@dlut.edu.cn)*

**Abstract**

In recent years, research on the interaction between Orbital Angular Momentum (OAM) and matter has seen a continuous influx of investigations. OAM possesses distinct properties, such as additional degrees of freedom, vortex characteristics, and topological properties, which expand its applications in optical communication, optical sensing, and optical force. Through experiments involving the interaction of a chiral metal swastika structure with a SAM-OAM beam generated by a q-plate, we have observed a phenomenon does not present in pure SAM beams. Fourier back focal plane (FBP) imaging under SAM beam excitation easily identifies the chirality and geometric properties of the structure. When the SAM-OAM beam excites the structure, FBP not only identifies its chirality and geometric properties but also distinguishes different OAM topological charges and signs, as well as the degree of elliptic polarization. The stokes parametric FBP imaging reveals asymmetric polarization distribution resulting from the interaction between a vortex beam and the chiral structure. Moreover, it clearly reflects the conversion process of SAM to OAM. The experimental results align well with simulation results. These findings hold valuable insights for the advancement of optical information storage and communication using OAM, opening up new possibilities for further exploration in this field.

**Keywords:** orbital angular momentum, chiral gammadion structure, Fourier back focal plane imaging, Stokes parameters.

**Introduction**

Vortices are widespread and fascinating phenomena in nature, observed in various contexts, from quantum vortices in liquid nitrogen and whirlpools in drains to massive events like hurricanes and spiral galaxies. These vortex phenomena are not confined to macroscopic matter alone; they also manifest in the realm of optics as optical vortices.[1] In contrast to circularly polarized beams that carry spin angular momentum (SAM), vortex beams with orbital angular momentum (OAM) have been extensively studied in optics since Allen first reported them in 1992.[2] Laguerre-Gaussian (LG) beams, known for their helical wavefront and straightforward experimental implementation, are extensively employed vortex beams with $\pm l\hbar$ OAM and a central phase singularity. Studies on this paraxially propagating vortex beam have revealed a new understanding of the intriguing connection between macroscopic optics and quantum effects. These investigations encompass various phenomena, such as twisting photons,[3-5] SAM-OAM interaction,[6-9] and Bose-Einstein condensates.[10-12] Due to their unique properties and advancements in optical manipulation technologies, vortex beams have been extensively researched in various fields, including optical communications, quantum memory, spin-orbit angular momentum conversion, light-matter interaction, optical tweezers, optical metasurfaces, super-resolution microscopy, vortex interferometers, sensing, and nonlinear optics.[13-29]

The inherent characteristics of OAM, enabling it to carry infinite angular momentum, render it an effective carrier for optical communication and optical memory. In the domain of light-matter interaction, the precise identification of OAM modes and nanostructures is of paramount importance. However, accomplishing accurate identification at the nanoscale level, comparable to the wavelength of light, remains a challenging task. The foundation of optical communication utilizing OAM lies in the capability to efficiently separate coaxially propagating beams with distinct OAM states. The identification of different OAM modes can be achieved at larger scales using conventional interferometry[30-31] and diffraction measurements.[32] In recent years, the recognition of OAM modes has been achieved in simulations and experiments at smaller scales as well. In 2012, Capasso et al. successfully integrated a designed interference pattern into a silicon photodiode based on the holographic principle, enabling selective detection of OAM.[33] In 2016, Qiu et al. utilized the constructive interference of helical-phase modulated surface waves excited by vortex beams to design a semi-ring plasmonic nanoslit, effectively achieving the discrimination of OAM.[34] In 2017, Kerber et al. proposed a method wherein the OAM information of light can be converted into spectral information through bright mode and dark mode, allowing direct readout of the additional information in twisted light using plasmonic nanoantennas.[35] In the same year, Miao et al. developed a simple and nondestructive method using plasmonic nanoholes to efficiently measure the OAM state of an optical beam.[36] In 2018, Zhu et al. demonstrated a technique using plasmonic gratings to spatially couple OAM modes into separate propagating surface plasmons with distinct splitting angles, allowing for the discrimination of OAM topological charge value and sign.[37] In two articles published by Kumar et al. in 2019[38] and 2022,[39] plasmonic nanowires were employed to identify SAM and OAM modes. However, the research remains confined to symmetric structures and has not yet been extended to more complex structures, such as asymmetric or chiral structures.

Indeed, the identification of the OAM beams with helical phases is closely intertwined with the SAM of light. The chiral media commonly employed in traditional SAM research can naturally be extended to facilitate the recognition and study of OAM.

Chirality is an intrinsic property commonly found in nature, exhibiting geometric properties of an object that do not coincide with its mirror image through simple rotation or translation.[40] In nature, chiral molecules exhibit weak optical activity (OA), but plasmonic metamaterials can significantly enhance the OA by amplifying the electromagnetic field[41] and generating super chiral field.[42-43] Various noble metal nanostructures, such as gammadion nanostructures, metal helices, overlapping metal strips, oligomers of nanodisks, and DNA-based assembled gold particles, are employed to achieve strong OA by enhancing the interaction between light and matter.[44-51] In recent years, OAM-OA has emerged as a captivating research field, driven by advancements in theory and experiments involving nanorods,[35, 52] chiral gammadion structures,[53-54] and dimensional chiral oligomers structures.[55] The above-mentioned chiral structures have primarily focused on OA, with limited studies on the imaging recognition of chiral structures.

Taking inspiration from the aforementioned research, we employ metal chiral gammadion structures to achieve rapid SAM and OAM chiral recognition using Fourier back focal plane (FBP) imaging technology. In this paper, we utilize the Laguerre-Gaussian beam generated by the q-plate to excite the gold gammadion structure, which is etched on the gold film using focused ion beam (FIB) technology. The reflected light is collected through an objective and passes through to be captured by a charge-coupled device (CCD) camera, facilitating Fourier images acquisition. By varying the SAM and OAM values, we can obtain distinct FBP images, enabling rapid identification and resolution of chiral structures' SAM and OAM characteristics. Additionally, parameter sweep studies are conducted for different sizes and elliptical polarization angles to further explore their effects on the imaging process. Through experimental data processing, the Stokes parameters distinctly reveal the SAM-OAM conversion, showcasing evident differences in the FBP images for the topological charges, structure chirality, and polarization. The numerical simulation results obtained using the finite element method (FEM) show excellent agreement with the experimental findings. These results hold significant potential for advancing the exploration of optical information storage and communication utilizing OAM.

**Results and discussion**

We have established a FBP imaging system capable of capturing both real images and Fourier images.[56] The sample used in our study is fabricated on a quartz wafer, coated with a $50\ nm$ thick gold layer using magnetron sputtering (Quorum-Q300T D plus). The gammadion structures are etched on the gold-coated wafer using a dual-focused ion beam system from Themo Fisher (FEI Helios G4 UX), employing a voltage of $30\ kV$ and a beam current of $7\ pA$. The schematic of the optical path is depicted in Figure 1. A $632.8\ nm$ He-Ne laser (REO-30995 from Excelitas Technologies) is directed through a linear polarizer (LP) and a quarter wave plate (QWP) to produce the necessary circularly polarized light (CPL). The beam then passes through a passive liquid crystal q-plate (Aroptix Switzerland, $q = 0.5$) to generate a vortex beam. This setup allows for precise control and generation of the desired optical vortex. After passing through a beam splitter, the beam is focused using a $50\ \times$ dark field objective, resulting in a focal spot with a diameter of approximately $4\ \mu m$ irradiated onto the sample surface. The reflected light is collected by the same objective and directed into a $4f$ system consisting of three lenses and a pinhole. This configuration effectively eliminates scattered light from other positions around the sample, enabling accurate and focused light collection for further analysis. By incorporating lenses with different focal lengths in front of the CCD camera (QImaging Retiga R1 from Cairn Research), we can effortlessly switch between obtaining real images and FBP images. The second QWP and LP are included in the setup to measure the potentially required Stokes parameters, providing comprehensive information about the polarization state of the light. This versatile experimental arrangement enables a comprehensive analysis of the chiral structures and their optical properties.

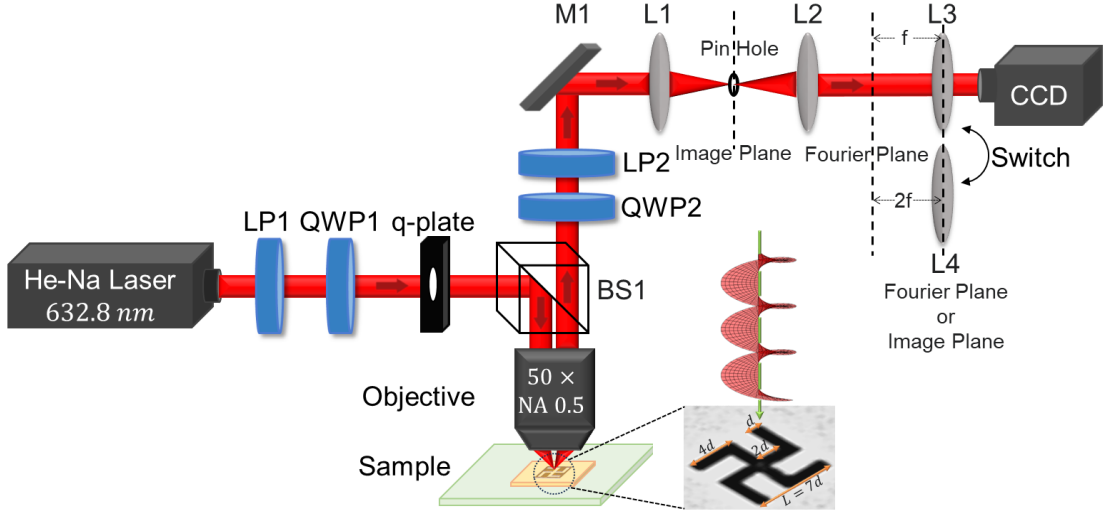

Figure 1. The schematic diagram of the FBP imaging optical path excited by the vortex beam. The setup includes LP1 and LP2 as linear polarizers, and QWP1 and QWP2 as quarter-wave plates. The first set of elements generates CPL, while the second set measures the Stokes parameters. BS1 serves as the beam splitter, and M1 functions as the mirror. L1, L2, and L3 represent lenses used in the $4f$ system. The sample illustration depicts the SEM image of the L-gammadion structure with a side length of $D = 4\ \mu m$ and $d = 714\ nm$.

By varying the angle between the QWP1 and the LP1, it is possible to generate elliptically polarized vortex light with varying ellipticities and azimuths. Figure 2a illustrates the intensity distribution of the spot of a real image, which was obtained using different vortex beams generated by the q-plate passing through the vertically polarized LP2. The diameter of the spot is approximately $4\ \mu m$. The upper right panel displays the polarization states of the vortex beams after the sign inversion of the q-plate, considering different incident polarizations. The LP1 is consistently set to vertically polarized, while $\theta$ ranges from $0°$ to $180°$, representing the angle between the fast axis of QWP1 and LP1. At $\theta = 0°$, the horizontal polarization lacks a horizontal component, resulting in a real image of the light spot passing through LP2 exhibiting a two-lobed pattern. As the angle $\theta$ increases, the pattern rotates by the same angle $\theta$, and the visibility of the crack line between the two lobes diminishes gradually. This is due to the gradual increase (albeit weak) in the horizontal component of the elliptical polarization at the crack region. As $\theta$ increases to $45°$, the pattern transforms into a ring with a small central hole. The uniformity of the ring spot indicates good quality. At each position of the spot, CPL predominates, resulting in almost equal vertical components. Consequently, the spot pattern after LP2 remains unchanged and is nearly symmetrical due to the balanced polarization properties at this angle. As $\theta$ increases from $0°$ to $180°$, the real image pattern of the intensity distribution of the spot undergoes periodic changes. The intensity distributions in the first and second rows appear very similar, but the rotation directions of the elliptical and circular polarizations between the two rows are opposite. Figure 2b depicts the FBP image corresponding to the real image of the light spot. For the $\theta = 0°$ ring pattern, the FBP image is horizontally cut, a distinctive feature of radial vortex beams. As $\theta$ increases, the rotation of the FBP image follows a similar pattern to the real image in Figure 2a. The separation at the crack line becomes increasingly blurred and eventually disappears. At $\theta = 45°$, the FBP image also transforms into a uniform ring spot, with the two lobes connected. The periodic changes observed in the real image with increasing $\theta$ also manifest in the FBP image. At this specific angle, the FBP image intensity distributions of $\theta = 45°$ and $\theta = 135°$ become indistinguishable. Consequently, the original vortex beam with opposite topological charge cannot be recognized by FBP without passing through the sample.

Indeed, vortex beams with varying angles can be represented equivalently by a first-order Poincaré sphere, as depicted in Figure 2c.[57-58] The representation of the polarization state involves using ellipticity $\varepsilon$ and azimuth $\alpha$. These parameters are employed to characterize the first-order vector beam, which is a polarized beam with spatially varying polarization states. The first-order vector beam is characterized by a constant ellipticity while exhibiting symmetrical rotation with the azimuth. The intersections of the equator and the S1 axis correspond to radial and azimuth vector beams, generated by vertical and horizontal linearly polarized light, respectively. Conversely, the intersections of the equator and the S2 axis represent circularly polarized vector beams, generated by $45°$ and $135°$ linearly polarized light, respectively. The first-order Poincaré sphere defines the vector beams generated by left-circularly polarized light (LCP, $s = -1$) and right-circularly polarized light (RCP, $s = +1$) with additional azimuthal phase and OAM of $\pm 1$, respectively. Any arbitrary vector beam, as shown in Figure 2c, corresponds to a specific point on the Poincaré sphere, with its latitude representing the ellipticity $2\varepsilon$ and longitude representing the azimuth $2\alpha$. By fitting the experimental parameters onto the Poincaré sphere, one can easily determine the conditions of the patterns observed in Figure 2a. Employing an analyzer, it becomes possible to discern the polarization states and other characteristics of the vector beams, aiding in the analysis and understanding of the experimental results.

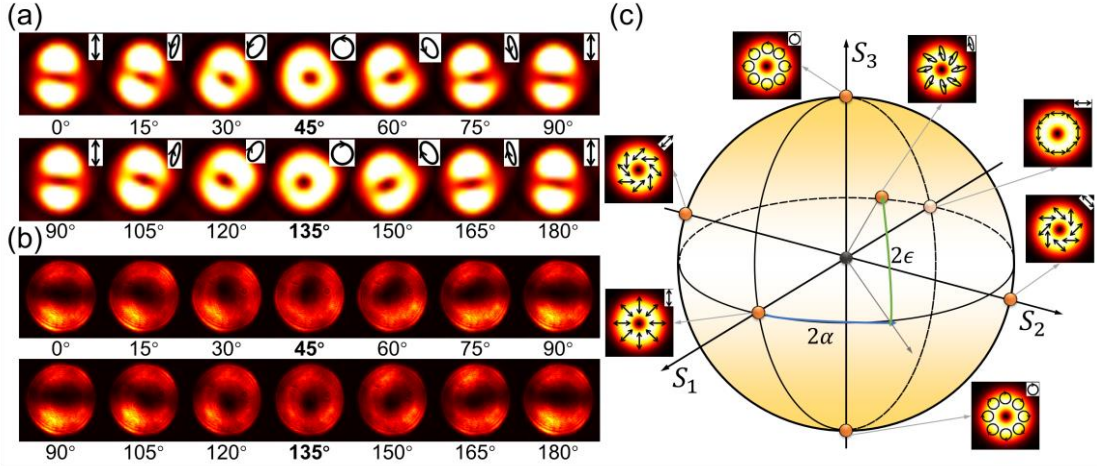

Figure 2. (a) The real image and (b) FBP image of the vortex beam spot in different polarization states. (c) Schematic representation of various vector beams on the first-order Poincaré sphere, including radial, azimuthal, circularly polarized vortex beams, and arbitrary state vortex beams represented by azimuth and ellipticity.

By substituting the blank substrate with chiral gammadion structures, we investigate the interaction between vortex beams carrying SAM and OAM with these structures. This interaction differs significantly from the traditional chiral light, which solely carries SAM. Figure 3 illustrates the distinct behavior and effects resulting from the combined SAM and OAM contributions in the presence of chiral gammadion structures. Figure 3a presents the scanning electron microscope (SEM) image of the samples utilized in the experiment. The left side of the panel showcases the hollow left-handed (L-) gammadion structures with side lengths $D$ ranging from 3 to 5 $\mu m$. On the right side, the image displays the cross structures with $D$ varying from 3.5 to 5 $\mu m$ (The SEM images of all samples are shown in Supplementary Information S1 Figure S1). Figure 3b illustrates a schematic diagram of the light spot used to excite the sample. The upper panel displays the circularly polarized vortex beams with $l = +1$, $s = -1$ and $l = -1$, $s = +1$, respectively. The lower panel represents CPL with $l = 0$, $s = +1$ and $l = 0$, $s = -1$, respectively. The direction of the arrow indicates the state of circular polarization. Figures 3c, d present the FBP images of the L-gammadion structure when excited by the SAM-OAM beam and the bare SAM beam, respectively. In the first row of Figure 3c, FBP images of the corresponding gammadion structures shown in the same column are excited by a vortex beam with $s = -1$ and $l = +1$. The FBP images reveal a transition of the spiral image from blurred to clear and, subsequently, to a near-cross pattern as the D varies from 3 to 5 $\mu m$. This phenomenon is directly influenced by the diameter of the vortex spot, which measures approximately 4 $\mu m$ after being focused by the objective. When structures smaller than 4 $\mu m$, effective excitation is hindered by the vortex beam's hollow characteristics. However, when the structure size matches the spot size, effective excitation occurs. As the size is increased beyond the spot size, the arms of the gammadion extend beyond the beam ring and lose efficient excitation, making it similar to a cross being excited. Consequently, the FBP image takes on a near-cross structure. The generation of a hollow FBP is attributed to the inherent phase singularity of the vortex beam. Indeed, the OAM helicity of the vortex beam is clockwise and opposite to the counterclockwise helicity of the SAM shown in the Figure 3b. Consequently, all FBP images undergo a certain clockwise rotation after passing through the space spiral and being reflected. This phenomenon arises from the inherent helical phase transmission characteristic of the vortex beam.

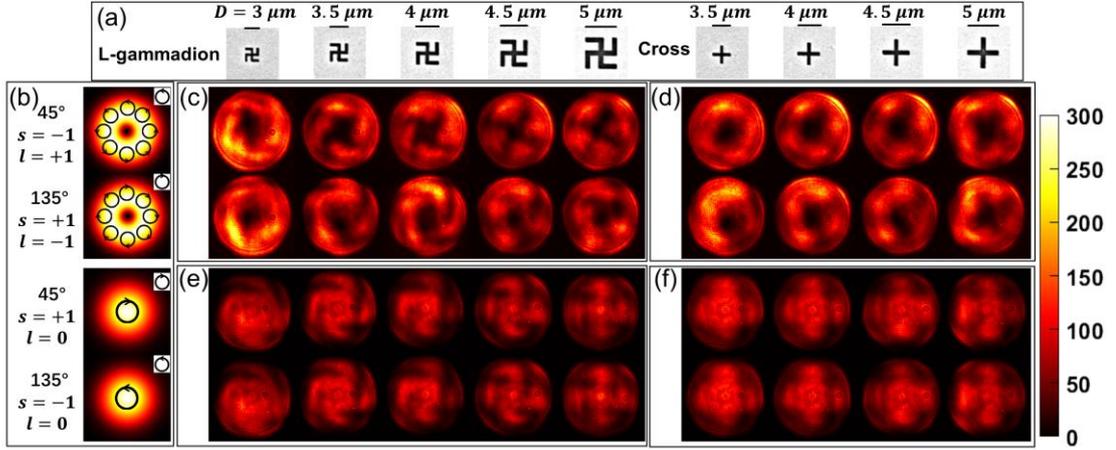

Figure 3. FBP images obtained from experiments involving hollow gold L-gammadion and cross structures excited using vortex beams and CPL. (a) SEM image showing the hollow L-gammadion (side length $D = 3$ to $5\,\mu m$) and cross structure ($D = 3.5$ to $5\,\mu m$). (b) Schematic diagrams of the vortex beam ($s = -1$ $l = +1$, and $= +1$ $l = -1$) and CPL ($s = +1$, and $s = -1$) are illustrated. (c-d) FBP images of the L-gammadion and cross structures excited by the vortex beam. (e-f) Corresponding FBP images excited by CPL.

The second row of Figure 3c illustrates the gammadion excited by an opposite chirality vortex beam with $s = +1$ and $l = -1$. It exhibits a similar phenomenon as described above, with the distinction being that the FBP image is rotated counterclockwise by a certain angle. This observation indicates that plasmonic nanostructures can effectively differentiate and respond to the rotation and chirality of the vortex beam. The FBP image of the symmetric cross structure excited by the vortex beam in Figure 3d shows FBP images with the same rotation angle. The response of the symmetric cross structure to the vortex beam's rotation effect upon reflection is not sufficient for distinguishing the beam uniquely. In contrast, when the vortex beam is scattered by the gammadion structure, the response not only changes the helical pattern but also results in different intensity distributions in the FBP images. This makes the FBP images of the gammadion structure distinct and allows for a clear differentiation between the incident vortex beams with different chirality and rotations. Subsequently, the opposite R-gammadion chiral structure exhibits a right-handed spiral FBP image, validating the experiment's capability for identifying vortex beams (Supplementary Information S2 Figure S2). In comparison, when the gammadion structures are exclusively excited by CPL as shown in Figure 3e, a solid left-handed spiral FBP image is observed. As the L-gammadion structure size increases, the FBP image evolves from blurred to clear and eventually transforms into a near cross shape. However, there is no clockwise rotation phenomenon in the FBP images of the LCP and RCP.

In the FBP images of the LCP and RCP, no clockwise rotation phenomenon is observed. As a result, the left and right rotations of the incident beam cannot be distinguished. The control experiment for the crosses excited by CPL is shown in Figure 3f, where a near cross shape similar to Figure 3e is observed. In summary, the vortex beam can recognize both the size and chirality of the gammadion structure, while the chiral structure exhibits a reverse recognition effect on the vortex beam with opposite OAM. This phenomenon is distinct from the behavior observed with the CPL. Additionally, the infinite topological charges of OAM hold the potential for novel optical phenomena.

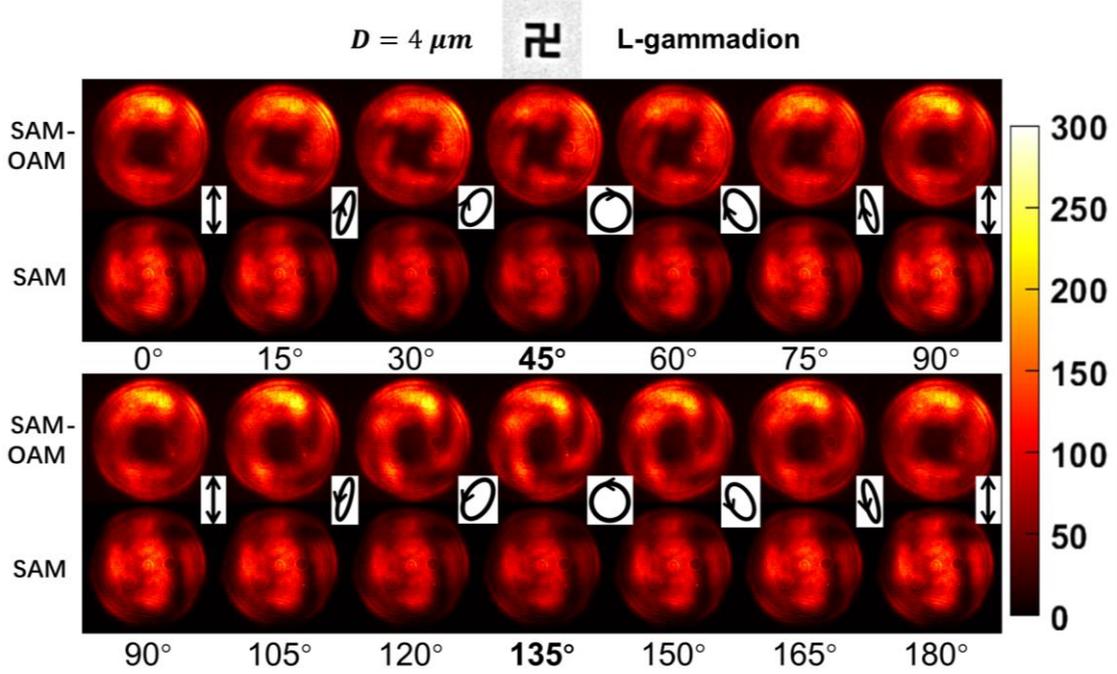

Figure 4. FBP images of the L-gammadion structure with $D = 4\,\mu m$ excited by CPL and SAM-OAM beam in various polarization states. The black arrow indicates the polarization state of the incident beam before passing through the q-plate.

For a detailed investigation of the interaction mechanism, we focused on the most efficient interaction size ($d = 4\,\mu m$) of the L-gammadion chiral structure. Various incident polarizations, corresponding to the angle between the LP1 and the QWP1 ranging from $0°$ to $180°$, are employed to generate vortex beams by passing through the q-plate. In Figure 2, the elliptically polarized light is transformed into a vortex beam with the opposite polarization state after passing through the q-plate. This resulting vortex beam is then used to excite the gammadion structure, as shown in Figure 4. The black arrow represents the incident polarization state before passing through the q-plate. It should be noted that the CPL does not pass through the q-plate. When the angle $\theta$ deviates significantly from $45°$ and $135°$ the beam gradually transforms into a radial vortex beam ($l = 0,\ s = \pm 1$). Consequently, the FBP displays an incomplete hollow spiral image due to the incident linear polarization characteristics. Exactly, when the incident beam is closer to CPL at $\theta = 45°$ ($l = +1, s = -1$), the beam transforms into a circularly polarized vortex beam. This configuration leads to a complete and clear spiral FBP image. The FBP pattern of the hollow shape rotates by approximately 45° compared to the sample hollow shape. Circularly polarized vortex beams of opposite chirality and helicity are generated at $\theta = 135°$. The FBP pattern of the hollow shape rotates by approximately $-45°$. The FBP pattern exhibits obvious differences with changing angles, periodically transitioning from blurred to clear. Additionally, the chiral structure has the potential to distinguish vortex light with different incident polarizations. With the change of the angle, the FBP pattern exhibits obvious differences, periodically transitioning from blurred to clear. The FBP patterns in the opposite spiral direction also show periodic transitions, indicating the chiral structure's potential to distinguish vortex beams with different incident polarizations. When the L-gammadion structure is excited with only the SAM beam, it shows insensitivity to the polarization state change, and the FBP at all angles does not exhibit any clockwise rotation. This observation further supports the influence of the introduced OAM helical space phase on the chiral structure. Additional details of the R-gammadion and cross structures are available in Supplementary Information S3 Figure S3, S4. The ability of the chiral structure to perceive the incident polarization of the vortex beam is verified. Exploring a wider range of polarization state changes and increasing topological charges may further enhance the structure's capability for identification and characterization.

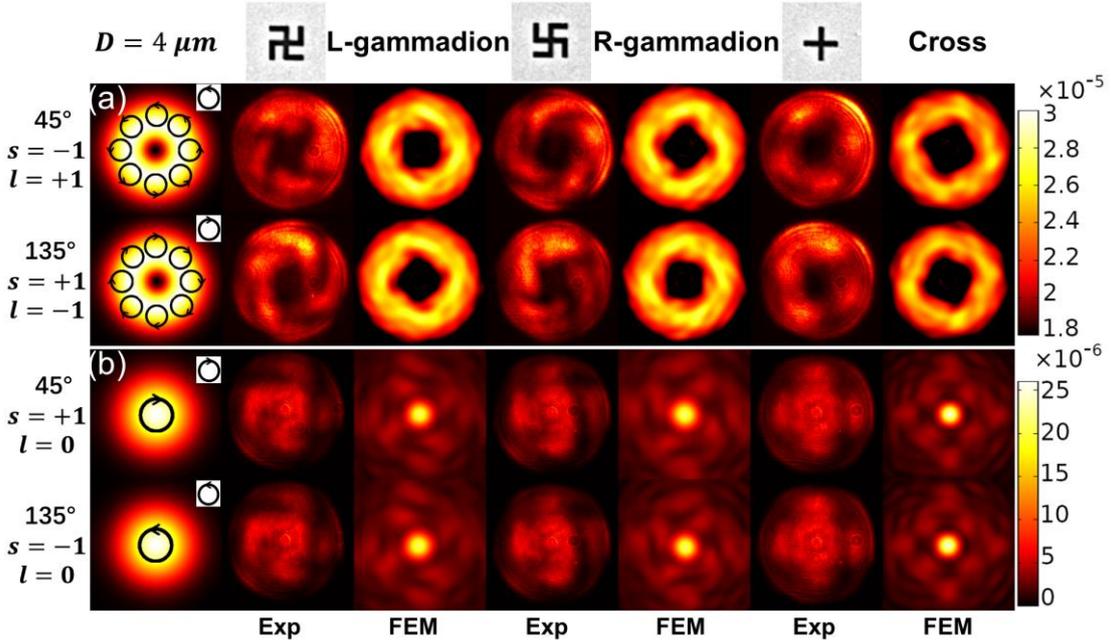

Figure 5. Comparison of far-field images of L- and R- gammadion and cross structures obtained through FEM numerical simulation and experimentally obtained FBP images. The structure size is $D = 4\ \mu m$, and the excitation light source is L-G beam and CPL at 45° and 135° ($l = \pm 1, s = \mp 1$ and $l = 0, s = \pm 1$). The color bar represents the FEM far-field range, and its absolute value is not important due to the background electric field being set to $1\ V/m$.

To gain deeper insights into the interaction phenomenon, numerical simulations were performed on the mentioned structure, which also served to validate the experimental results. COMSOL Multiphysics 5.5 was utilized for simulating hollow gammadion structures of various sizes. The far-field data obtained from the simulations were then processed to obtain the results presented in Figure 5. In this section, we present a comparison between experimental and simulation results of the FBP image for the L-/R-gammadion and cross structures (with the same size as shown in Figure 4) under illumination by vortex beams ($l = \pm 1, s = \mp 1$) in Figure 5a and CPL ($l = 0, s = \pm 1$) in Figure 5b. The light source utilized is an L-G beam with a beam waist radius of $2\ \mu m$, which matches the experimental setup. The Perfect Matching Layer (PML) used in the simulation has a thickness of $400\ nm$, and the maximum mesh size is set to be less than $100\ nm$ to ensure its fine resolution. The dielectric constant of gold is taken from experimental data reported by Johnson and Christy.[59] In the simulations, similar rotating hollow spiral FBP images were obtained, replicating the experimental results. The FBP vortex patterns exhibited the same left/right-handed rotation direction as the sample structure. Specifically, the L-gammadion structure excited by ($l = +1, s = -1$) and the R-gammadion structure excited by ($l = -1, s = +1$) displayed the same patterns, and vice versa. For the cross structure, there is no significant difference in the FBP images, except for a rotational angle. This observation is further supported by simulation, indicating that OAM can indeed play a role in identifying chiral structures. Furthermore, the introduction of the metal structure disrupts the symmetrical intensity distribution of $l = +1$ and $l = -1$ vortex beams, enabling the identification of OAM beams with opposite topological charges through differences in the FBP intensity distribution. When the same structure is excited by the CPL of opposite SAM, it exhibits only a symmetric pattern for $s = +1$ and $s = -1$. This indicates the difficulty of the structure in identifying the SAM through the FBP intensity distribution.

To further investigate the mutual recognition capabilities of vortex beams and chiral structures, we employed Stokes parametric FBP to study the optical properties of vortex beams in various polarization states and their interactions with chiral structures. We performed the experiment and obtained the Stokes parametric FBP images with $S_0$ to $S_3$ by adjusting the QWP2 and LP2 in the detection path. Stokes parametric analysis provides a more effective description of the polarization state of electromagnetic radiation. In this study, we specifically focus on the Stokes parametric FBP of the R-gammadion structure with $D = 5\ \mu m$ under excitation by CPL and OAM beams, respectively.

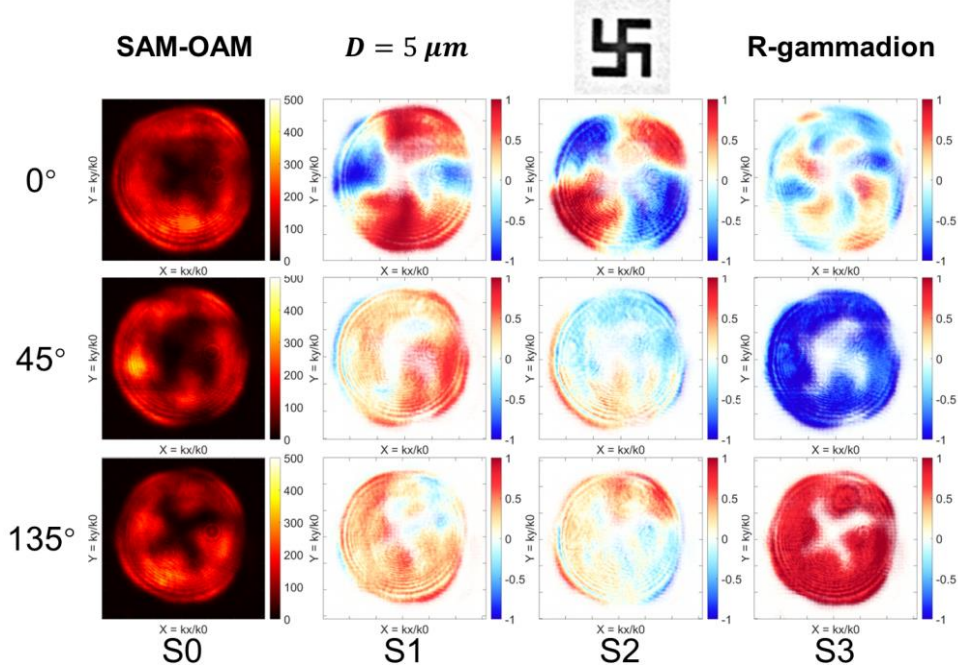

Figure 6. Stokes parametric FBP images of R-gammadion structure excited by different types of vortex beams. The side length of the R-gammadion structure is $D = 5\ \mu m$.

As depicted in Figure 6, the first row corresponds to a radial vortex beam, explaining the four-petal distribution of $S_1$ and $S_2$, which is expected.[60] In conventional cases, the value of $S_3$ should be 0. However, in this scenario, $S_3$ shows a unique behavior with a positive and negative spiral pattern, which is not observed in $S_0$. In contrast, the relative strength of the positive and negative components in the cross structure is only half of that observed in the chiral structure (refer to the Supplementary Information S4 Figure S5). The stronger interaction of the radial vortex beam with the chiral structure results in an asymmetric circular polarization distribution. This asymmetry is attributed to the introduction of the chiral asymmetric geometry in the structure, further influencing the response to the incident vortex beam. Moreover, for the 45° and 135° incident circularly polarized vortex beams, the $S_3$ parameter exhibits a completely negative and positive distribution, respectively, which is consistent with expectations. The spiral pattern of $S_3$ is almost identical to that of $S_0$, which is clearly distinct from the above radial vortex beam. However, the observed small values of $S_1$ and $S_2$, different from the expected value of 0. This non-zero linear polarization component is a result of the superposition of asymmetrical changes in the circular polarization component caused by the interaction with the chiral structure. The similarity between the $S_1$ and $S_2$ distributions of the cross structure and the previous observations is expected, as both structures are likely to exhibit comparable responses to linear polarization. The variation in the Stokes parametric FBP images of the gammadion structure when excited by the OAM beam confirms that their interaction induces a change in the polarization state. This observation supports the ability of the chiral structure to distinguish vortex beams with different OAM values.

Under linearly polarized light and CPL excitation (Supplementary Information S4

Figure S6), the Stokes parametric FBP images reveal that the $S_1$ parameter of 0° linear polarization is positive, while $S_2$ and $S_3$ show very weak intensity which is in line with expectations. However, for CPL excitation, $S_1$ and $S_2$ exhibit positive and negative values, respectively, but are much weaker compared to the relative intensity of the vortex beam. We have reason to believe that the interaction between circularly polarized vortex beam and chiral structures results in more asymmetric responses compared to LCP or RCP illumination. The opposite sign of the value compared to that of the vortex beam provides further evidence of the successful chiral inversion of the q-plate for the incident CPL. The stronger $S_3$ value of the vortex beam compared to that of the CPL indicates a higher circular polarization asymmetry in their interaction, making it more conducive to the chiral structure's ability to recognize the OAM beam.

**Conclusion**

In this study, we perform reflectance FBP imaging of the gold chiral gammadion structure under vortex beam and CPL excitation. We investigate the influence of parameters such as incident polarization, gammadion structure size, and chiral value on their interaction. At the same time, the FBP imaging of gammadion structures illuminated by vortex beam compared to only CPL exhibits distinguishable phenomenon. The inherent spatial phase characteristic of the vortex beam enables the FBP image to distinguish between OAM and SAM based on its clockwise rotation. Additionally, it can be utilized to differentiate various OAM topological charges. The results obtained from FEM numerical simulations confirm the reliability of the experimental findings. Furthermore, the Stokes parametric FBP images provide valuable insights into the polarization distribution during the interaction, highlighting the effective interaction between the vortex beam and the chiral structure. The opposite sign of $S_3$ for the vortex beam with $l = \pm 1, s = \mp 1$ and the spin beam with $l = 0$, $s = \pm 1$ confirms the SAM-to-OAM conversion. Overall, the vortex beam demonstrates its ability to recognize the chiral metal structure and exhibits sensitivity to both the structure's size and chirality. Furthermore, the chiral structure can distinguish between OAM and SAM, discern OAM with opposite topological charge, and display sensitivity to incident polarization. The SAM-OAM beam used in this study represents just one type of vortex beam. We hope these findings will inspire future investigations to identify additional topological charges and explore diverse vortex beam patterns.

**Funding**
National Natural Science Foundation of China (NSFC) (12074054, 12274054).

**Conflicts of interest**
The authors declare no competing financial interest.

**Data Availability**

The data that support the findings of this study are available from the corresponding author upon reasonable request.

**Supporting Information**

The Supporting Information is available on the website at DOI:xxx-xxx, including the SEM images of all samples (Figure S1), FBP images of hollow gold R-gammadion structures (Figure S2 and S3) and cross structures (Figure S4), Stokes parametric FBP images of gammadion and cross structures (Figure S5 and S6).

# Distinguishing the topological charge of vortex beam by Fourier Back Plane Imaging with Chiral Gammadion Structure

Yangzhe Guo[1], Jing Li[2] and Yurui Fang[1,*]

[1.] School of Physics, Dalian University of Technology, Dalian 116024, P.R. China.

[2.] Key Laboratory of Photochemical Conversion and Optoelectronic Materials, Technical Institute of Physics and Chemistry, Chinese Academy of Sciences, Beijing 100190, China

*Corresponding author: Yurui Fang (yrfang@dlut.edu.cn)

**S1. The SEM images of all samples.**

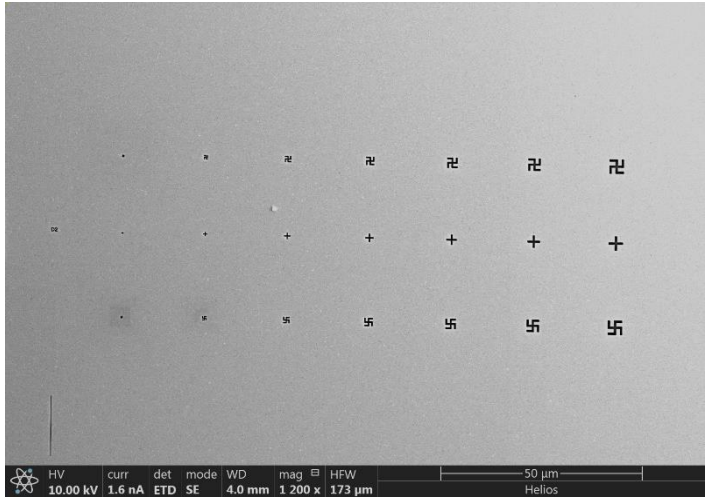

Figure S1. SEM images of all hollow L/R-gammadion and cross structures etched on $50\ nm$ gold film ($D = 2$ to $5\ \mu m$).

**S2. FBP images collected of hollow gold R-gammadion structures.**

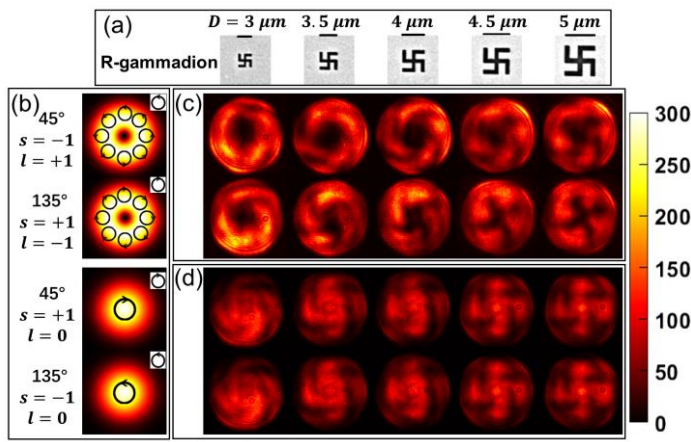

Figure S2. FBP images collected of hollow gold R-gammadion structures excited by vortex beam and circularly polarized light. (a) SEM image of R-gammadion (side length $3\ \mu m$ to $5\ \mu m$). (b) Schematic diagrams of vortex beam ($s = -1\ \ l = +1$, and $s = +1\ \ l = -1$) and circularly polarized light ($s = +1$, and $s = -1$). L-gammadion structures of FBP excited by vortex beam (c) and circularly polarized light (d).

## S3. FBP images of the R-gammadion and cross structures excited by CPL and OAM beam in various polarization states.

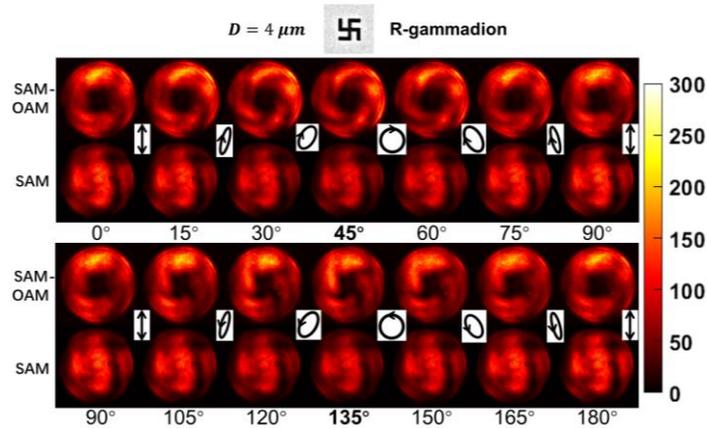

Figure S3. FBP images of the R-gammadion structure with $D = 4\ \mu m$ excited by CPL and SAM-OAM beam in various polarization states. The black arrow indicates the polarization state of the incident beam before passing through the q-plate.

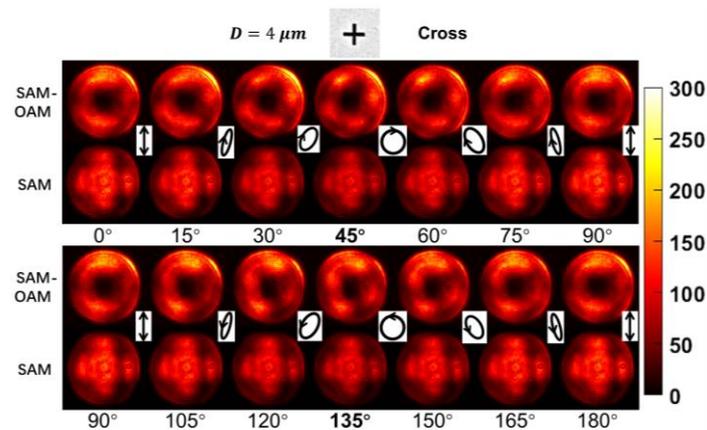

Figure S4. FBP images of the cross structure with $D = 4\ \mu m$ excited by CPL and SAM-OAM beam in various polarization states. The black arrow indicates the polarization state of the incident beam before passing through the q-plate.

## S4 Stokes parametric FBP images of gammadion and cross structures excited by vortex beam and CPL.

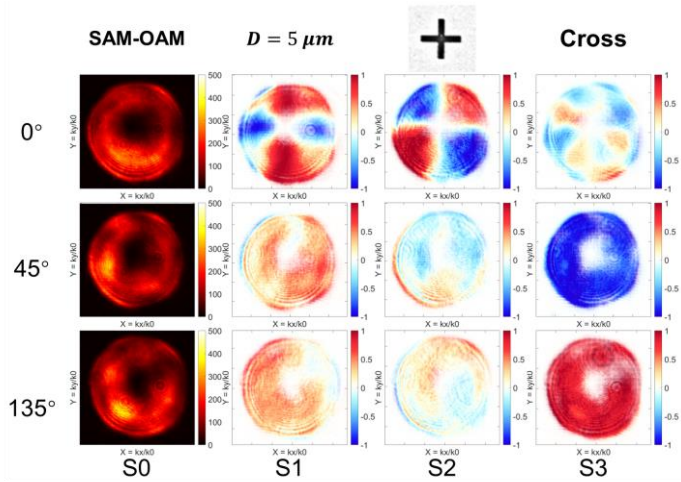

Figure S5. Stokes parametric FBP images of cross structure excited by different types of vortex beams. The side length of the cross structure is $D = 5\ \mu m$.

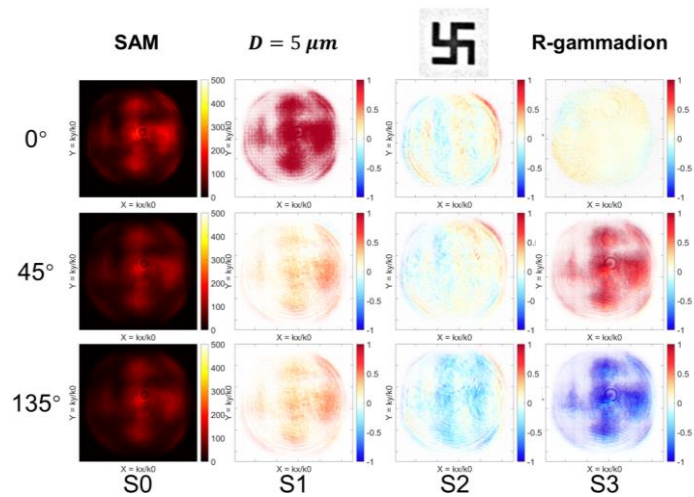

Figure S6. Stokes parametric FBP images of R-gammadion structure excited by different types of CPL. The side length of the R-gammadion structure is $D = 5\ \mu m$.